\documentclass[letterpaper,twocolumn,10pt]{article}
\usepackage{usenix}
\usepackage{times,datetime,url,hyperref,graphicx,multirow}
\usepackage{color,calc,ulem,threeparttable,tabularx,booktabs}
\usepackage{enumitem,comment,balance,subcaption,morefloats,mathrsfs}
\usepackage{algpseudocode}
\usepackage[group-separator={,}]{siunitx}
\usepackage{caption}
\usepackage{mathtools}
\usepackage{multirow}

\usepackage{listings}
\usepackage{multicol}

\usepackage{anyfontsize}

\definecolor{mygreen}{rgb}{0,0.6,0}
\definecolor{mygray}{rgb}{0.5,0.5,0.5}
\definecolor{mymauve}{rgb}{0.58,0,0.82}

\newcommand\figwidth{0.43\textwidth}

\usepackage{xcolor}

\colorlet{punct}{red!60!black}
\definecolor{background}{HTML}{EEEEEE}
\definecolor{delim}{RGB}{20,105,176}
\colorlet{numb}{magenta!60!black}

\lstdefinelanguage{json}{    literate=
     *{0}{{{\color{numb}0}}}{1}
      {1}{{{\color{numb}1}}}{1}
      {2}{{{\color{numb}2}}}{1}
      {3}{{{\color{numb}3}}}{1}
      {4}{{{\color{numb}4}}}{1}
      {5}{{{\color{numb}5}}}{1}
      {6}{{{\color{numb}6}}}{1}
      {7}{{{\color{numb}7}}}{1}
      {8}{{{\color{numb}8}}}{1}
      {9}{{{\color{numb}9}}}{1}
      {:}{{{\color{punct}{:}}}}{1}
      {,}{{{\color{punct}{,}}}}{1}
      {\{}{{{\color{delim}{\{}}}}{1}
      {\}}{{{\color{delim}{\}}}}}{1}
      {[}{{{\color{delim}{[}}}}{1}
      {]}{{{\color{delim}{]}}}}{1},
}

\lstdefinelanguage{Uppaal}{  keywords={after update,assign,before update,break,case,const,continue,
    default,else,enum,for,guard,if,meta,process,progress,return,select,
  state,transitions,sync,switch,trans,system,while},
  keywords={[2]broadcast,bool,clock,chan,commit,init,int,scalar,struct,
  typedef,urgent,void},
  keywordstyle={[2]\bfseries},
  keywords={[3]false,true},
  otherkeywords={[3]−>},
  morekeywords={[3]−>},
  keywordstyle={[3]\bfseries},
  comment=[l]{//},
  morecomment=[s]{/∗}{∗/},   commentstyle=\itshape,     captionpos=b,   escapechar=@, }

\lstdefinelanguage[GUI]{Uppaal}[]{Uppaal}{  keywordstyle={[2]\color{black!50!green}},   otherkeywords={−>},
  keywordstyle={[3]\color{magenta}},
  commentstyle={\color{black!50!red}\itshape},   literate={{−−>}{$−−>$}3} }

\usepackage{algorithm}
\usepackage{algorithmicx}

\algnewcommand{\LineComment}[1]{\State{\(\triangleright\) #1}}

\algnewcommand\algorithmicswitch{\textbf{switch}}
\algnewcommand\algorithmiccase{\textbf{case}}
\algnewcommand\algorithmicassert{\textbf{assert}}

\algnewcommand\Assert[1]{\State \algorithmicassert(#1)}
\algdef{SE}[SWITCH]{Switch}{EndSwitch}[1]{\algorithmicswitch\ #1\ \algorithmicdo}{\algorithmicend\ \algorithmicswitch}\algdef{SE}[CASE]{Case}{EndCase}[1]{\algorithmiccase\ #1}{\algorithmicend\ \algorithmiccase}\algtext*{EndSwitch}\algtext*{EndCase}
\usepackage{bm}

\hypersetup{  bookmarks=true,
  unicode=true,
  pdftoolbar=true,
  pdfmenubar=true,
  pdffitwindow=true,
  pdfstartview={FitV},
  pdfnewwindow=true,
  colorlinks=false,
  pdfdisplaydoctitle=true,
  pdfborder={0 0 0}
}
\usepackage[all]{hypcap}

\usepackage[absolute]{textpos}
\textblockorigin{0.75in}{0.875in}

\setlist[itemize]{leftmargin=*,partopsep=5pt}
\setlist[enumerate]{leftmargin=*,partopsep=5pt}

\usepackage{dcolumn}
\newcolumntype{d}[1]{D{.}{.}{#1}}

\newcommand\nil{\textit{nil}}

\usepackage{tikz}

\makeatletter
\newcommand*{\rome}[1]{\expandafter\@slowromancap\romannumeral #1@}
\makeatother

\newcommand{\verifi}{\textsc{VeriFi}}
\newcommand{\jammer}{\textsc{PacketSniper}}
\newcommand{\wifi}{Wi-Fi}
\newcommand{\us}{$\mu s$}

\newcommand{\m}[1]{m_{#1}}

\newcommand{\authreq}{\texttt{AUTH\_REQ}}
\newcommand{\authresp}{\texttt{AUTH\_RESP}}
\newcommand{\assocreq}{\texttt{ASSOC\_REQ}}
\newcommand{\assocresp}{\texttt{ASSOC\_RESP}}
\newcommand{\clientack}{\texttt{CLIENT\_ACK}}
\newcommand{\apack}{\texttt{AP\_ACK}}

\newcommand{\verifier}{\textsc{TraceVerifier}}
\newcommand{\analyzer}{\textsc{ProtocolAnalyzer}}
\newcommand{\uppaal}{\textsc{Uppaal}}
\newcommand{\rv}{Shi \textit{et al.}}
\newcommand{\lstline}[1]{L\ref{#1}}
\newcommand{\tx}{\texttt{T}}
\newcommand{\rx}{\texttt{R}}

\makeatletter

\global\def\section{\@startsection {section}{1}{\z@}                                   {2ex \@plus 1ex \@minus .1ex}                                   {1ex \@plus.2ex}                                   {\normalfont\bfseries\scshape\fontsize{11}{13}\selectfont}}
\global\def\subsection{\@startsection{subsection}{2}{\z@}                                     {2ex\@plus 1ex \@minus .1ex}                                     {1ex \@plus .2ex}                                     {\normalfont\bfseries\fontsize{10}{12}\selectfont}}
\global\def\subsubsection{\@startsection{subsubsection}{3}{\z@}                                     {2ex\@plus 1ex \@minus .1ex}                                     {1ex \@plus .2ex}                                     {\normalfont\itshape\fontsize{10}{12}\selectfont}}

\global\def\@maketitle{  \newpage
  \begin{center}  \let \footnote \thanks
  \null
    \vskip -.3em    {\bf\LARGE \@title \par}    \vskip 1em    {\large
      \lineskip .5em      \begin{tabular}[t]{c}        \@author
      \end{tabular}\par}    \vskip 1em    {\large \@date}  \end{center}  \par
  \vskip -0.2em
	}

\def\theconference{NSDI'18}
\def\thetitle{VeriFi: Model-Driven Runtime Verification Framework for Wireless Protocol Implementations}
\def\theauthors{}

\ifdefined\isdraft
  \usepackage{fancyhdr}
\else
\fi

\begin{document}

\title{  \thetitle
}
  \date{}

\author{Jinghao Shi\\
University at Buffalo\\
\texttt{jinghaos@buffalo.edu}
\and
Shuvendu Lahiri, Ranveer Chandra\\
Microsoft Research\\
\texttt{\{shuvendu,ranvver\}@microsoft.com}
\and
Geoffrey Challen\\
University of Illinois at Urbana-Champaign\\
\texttt{challen@illinois.edu}
}

\hypersetup{  pdfinfo={    Title={\thetitle},
    Author={\theauthors},
  }
}

\newcommand{\shuvendu}[1]{{\bf[SL:{#1}]}}
\newcommand{\jiansong}[1]{{\bf[Jiansong:{#1}]}}
\newcommand{\gwa}[1]{{\textcolor{red}{\bf{GWA: #1}}}}
\maketitle

\begin{abstract}
		Validating wireless protocol implementations is challenging.
		Today's approaches require labor-intensive experimental setup and manual
	trace investigation, but produce poor coverage and inaccurate and
	irreproducible results.
    We present \verifi{}, the first systematic sniffer-based, model-guided runtime verification framework for wireless protocol implementations.
		\verifi{} takes a formal model of the protocol being verified as input.
		To achieve good coverage, it first applies state reachability analysis by
	applying model checking techniques.
		It then uses a new \jammer{} component to selectively trigger packet losses
	required to quickly investigate all reachable protocol states.
		Our results show that the selective packet jamming allows \verifi{} to significantly improve the coverage of protocol states.
		Finally, \verifi{} accommodates uncertainty caused by the sniffer when validating
	traces, allowing it to provide accurate and reproducible results.
		By modeling uncertainty, \verifi{} highlights likely protocol violations for
	developers to examine.
	\end{abstract}
 \section{Introduction}
\label{sec:intro}

The rapid increase in the number of mobile and custom wireless devices is
making wireless protocols increasingly important.
Wireless protocols specifications undergo extensive testing, typically through
simulations.
But a lack of good tools results in protocol \textit{implementations} not
being tested as carefully as protocol specifications.
Implementations can always introduce bugs, particularly when development is
done by parties that did not author the specification.
Given the rise of custom wireless devices---such as Apple TVs, XBox
controllers, ChromeCasts, and FitBits---effective testing of wireless
protocol implementations is more important today than ever before.
These devices frequently implement custom and sometimes proprietary wireless
protocols.
The implementation may itself be proprietary, making source code unavailable
to the party performing validation.
And they may lack capabilities required to do white-box testing.
A series of connectivity flaws in deployed devices and systems---including
Apple Watch 3~\cite{app_watch}, iOS~8~\cite{wifried}, Google Android Lollipop~\cite{lollipop}, and the
Microsoft Surface Pro 3~\cite{surface}---help demonstrate that a new approach
to wireless protocol implementation verification is needed.

Formal model checking techniques have been used to verify the correctness of
wired communication protocol
implementations~\cite{musuvathi2002cmc,engler2004model} and distributed
systems~\cite{feamster2005detecting,canini2012nice,liu2008d3s}.
But there are several unique features that make verifying wireless protocol
implementations more challenging.
Wireless protocols confront a more complex and dynamic environment than their
wired counterparts.
Environment variables---such as attenuation, multi-path, fading, and
interference---are hard to control, making reproducible experimentation
difficult.
To meet tight timing constraints, wireless protocols are often implemented in
low level firmware.
This makes it difficult to apply source code model checking techniques.
Finally, due to closed-source implementations, validation often must be done
using external wireless sniffers.
Due to the physical nature of the wireless medium, sniffers introduce
uncertainty that can sabotage the validation results.

The result of these difficulties is that no systematic testing or verification system for
wireless protocol implementations currently exists.
Current state-of-the-art industry practices begins with labor-intensive
conducted test setup.
Manual spot-checking is then used to inspect packet traces collected by the
sniffer and validate implementation correctness.
The process is tedious, error-prone and time-consuming, but also misses many protocol
states resulting in poor test coverage.

In this paper, we describe the design and implementation of \verifi{}:
the first systematic sniffer-based, model-guided runtime verification framework for wireless protocol implementations.
While \verifi{} could be extended to study timing-related failures, it
currently focuses on implementation bugs caused by packet loss.
Factors such as interference and fading make packet loss normal in wireless
communications.
A core task of wireless protocols is handling packet loss, and \verifi{} can
help validate that they do so correctly.

\verifi{} requires both a \textit{model} of the protocol specification and
access to the devices being tested.
The model consists of state machines for portions of the protocol that require
verification, such as {\it association}, {\it rate control}, and {\it retransmission} policies.
During testing, \verifi{} manipulates communication with the tested device to
ensure that it quickly reaches all reachable protocol states.
It accomplishes this using two novel components.
A {\em protocol model analyzer} determines packet loss sequences that can
drive the protocol model into various states.
The \jammer{} uses these sequences to determine if a packet currently in the
air should be lost to reach an untested state.
If so, it prevents packet reception by generating a jamming signal.

Once testing is complete, a {\em trace verifier} analyzes recorded packet
exchanges to identify protocol violations.
\verifi{} can distinguish likely implementation bugs from false positions that
were probably caused by sniffer uncertainty.
This produces a clear work flow for helping developers bring their
implementation in line with the specification.

Our work makes the following contributions:
\begin{itemize}
			\item To the best of our knowledge, \verifi{} is the first automatic and systematic framework for end-to-end validation of wireless protocol implementations.
			\item By formally describing the model of wireless protocols, we use model
		checking techniques to automatically infer the packet reception and loss
		sequences that drive the model and implementation into target states.
			\item We describe the design principles and two prototype implementations of
		\jammer{} for \wifi{} systems, which selectively drop the packets
		specified by the protocol model analyzer.
			\item We propose a new method of validating sniffer traces under uncertainty
		(as observed in~\cite{shi-rv16}).
				By incorporating sniffer as part of our model, the sniffer trace
		uncertainty can be tackled by an off-the-shelf model checker (instead of a specialized checker~\cite{shi-rv16}).
				\item We perform an end-to-end evaluation of our framework using the
			802.11 link setup protocol.
				It demonstrates that \verifi{} significantly improves test coverage,
		efficiency, and reproducibility.
				We also report three implementation issues \verifi{} discovered that do
		not manifest otherwise.
		\end{itemize}

While we believe that \verifi{} is a general approach that can validate many
different wireless protocol implementations, this paper focuses on Wi-Fi based systems.
 \section{Validation Framework}
\label{sec:system}

We use the \wifi{} link setup protocol as a concrete example to illustrate the
work flow of our validation framework. The detailed protocol is explained in
Section~\ref{subsec:protocol}. We then describe three key observations from the
example protocol in Section~\ref{subsec:observ} and show an overview of our
validation framework in Section~\ref{subsec:overview}.

\subsection{Example Protocol}
\label{subsec:protocol}

\begin{figure}[t!]
  \centering
  \begin{subfigure}{\figwidth{}}
    \includegraphics[width=\textwidth]{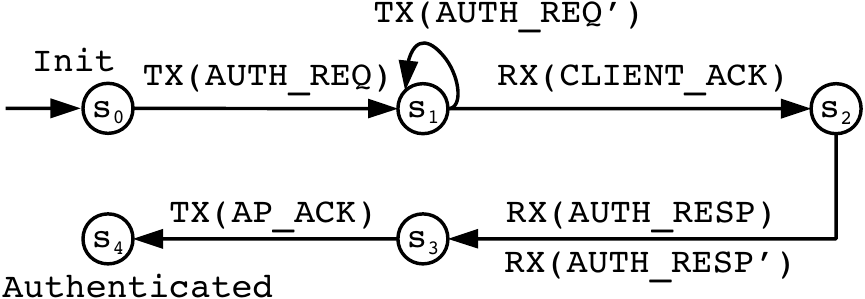}
    \caption{\textbf{Client State Machine for the Authentication Step.} \texttt{PKT'}
      denotes the MAC layer retransmissions of \texttt{PKT}.}
    \label{fig:assoc_sm:client}
  \end{subfigure}\\\vspace*{4mm}
  \begin{subfigure}{\figwidth{}}
    \includegraphics[width=\textwidth]{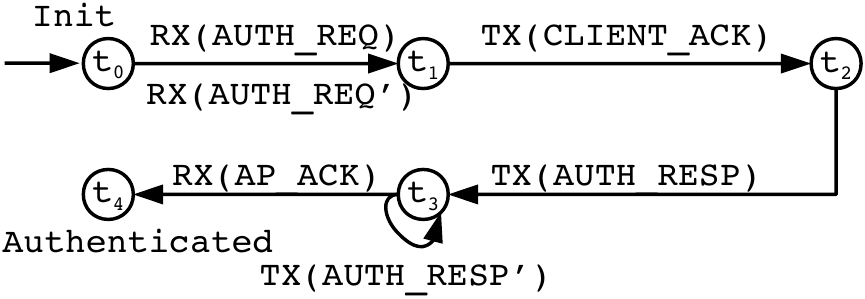}
    \caption{\textbf{AP State Machine for the Authentication Step.}}
    \label{fig:assoc_sm:ap}
  \end{subfigure}
  \caption{\textbf{802.11 Link Setup Protocol and State Machines.}}
  \label{fig:assoc_sm}
\end{figure}

The \wifi{} link setup protocol consists of three phases: authentication,
association and 802.11X authentication. Figure~\ref{fig:assoc_sm:client} shows
the client's state machine for the authentication phase. The client first sends
\authreq{} and expects to receive \clientack{} ($s_0\rightarrow s_1$). Either it
receives \clientack{} and then waits for \authresp{} ($s_1\rightarrow s_2$), or
it retransmits \authreq{} and remains at $s_1$.  Once it receives both the
\clientack{} and the \authresp{} packet, it moves to the \texttt{Authenticated}
state $s_4$. The client then continues to start the association request/response
handshake process, the detailed state machines of which are omitted for sake of
space.

In Figure~\ref{fig:assoc_sm:ap}, when the AP receives \authreq{} or
its retransmission, it first acknowledges the request ($t_1\rightarrow t_2$),
and then replies with \authresp{}. Similarly, multiple
retransmissions of \authresp{} may occur before the AP receives \apack{},
and is ready to handle the association request from this
client. The AP may have several such state machines running in parallel to
handle the authentication requests from multiple clients.

\subsection{Key Observations}
\label{subsec:observ}

We make several observations from the example protocol described in
Section~\ref{subsec:protocol}. First, wireless protocols can be modeled as a
collection of state machines that interoperate with each other via packet
exchanges. In particular, each end device can be treated as a black box and the
inputs of the state machine are limited to only externally observable events:
what (and when) packet is transmitted. This constraint not only simplifies the
protocol analysis and verification, but also has practical benefits. As
explained earlier, the implementations of wireless protocols are often
proprietary, making observing end-points' internal states extremely difficult if
not entirely impossible. Certain wireless protocol aspects, such
as the Clean Channel Assessment (CCA) and Distributed Coordination Function
(DCF), cannot be modeled in this manner as they require access to the end
device's internal states.

Second, packet loss is the key factor that alters the end-point behaviors in
wireless protocols. Other factors, such as out-of-order delivery and queueing
delay, are less significant in wireless protocols than their wired counterparts.
For instance, the link setup protocol in Figure~\ref{fig:assoc_sm} becomes
trivial in the absence of packet loss: each device just iterates through the
three stages sequentially before both end-device reach the same associated
state.  In fact, we argue that a large portion of any wireless protocols is to
define how the endpoints cope of packet losses: retransmissions and
acknowledgments are required to ensure packet delivery, rate control mechanisms
are used to improve the link performance in face of packet loss, etc.

Finally, validating wireless protocol implementations involves driving the
system to certain state, and then observe and validate the Device Under Test
(DUT)'s behavior. The system state can be represented by a tuple of the model
state at each endpoint. For instance, in the state machines shown in
Figure~\ref{fig:assoc_sm:client} and~\ref{fig:assoc_sm:ap}, the system state
$\langle Client.s_1, AP.t_3 \rangle$ represents the case where the client and
the AP disagree with the authentication status: the client thinks it is not
authenticated and is trying to retransmit the \authreq{} packet, while the AP
actually receives the \authreq{} packet and is trying to transmit the
\authresp{} packet.

\subsection{Framework Overview}
\label{subsec:overview}

\begin{figure}[t!]
  \centering
  \includegraphics[width=\figwidth{}]{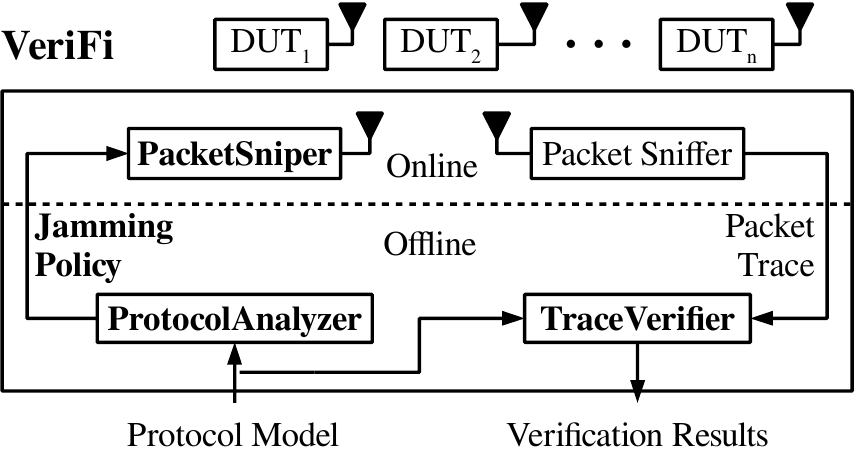}
  \caption{\textbf{\verifi{} Framework for Verifying Wireless Protocol Implementations.}
  Components in bold face are proposed and described by this paper. DUT stands
for Device Under Test.}
  \label{fig:system}
\end{figure}

Based on the observations, we propose \verifi{}, a model-driven runtime
verification framework for wireless protocol implementations.
Figure~\ref{fig:system} shows main components and the work flow of the
framework.

Beginning  with a formal model of the protocol, the \analyzer{}
first uses model checking techniques to infer the edge sequence (packet
success/loss) to reach each system state. We describe general principles of
modeling wireless protocols and one specific realization using \uppaal{} model
checker in Section~\ref{sec:policy}.

The packet success/loss information is then fed to \jammer{}---a real-time
reactive packet jammer. The \jammer{} drops the packets
that are tagged as loss in the policy. We describe the design and two
implementations of \jammer{} for 802.11 in Section~\ref{sec:jammer}.

Finally, packet traces are collected to verify the system's behavior by the
\verifier{} component. As observed in~\cite{shi-rv16}, sniffer trace can not
directly be used for verification due to uncertainty. We describe how
\verifi{} handles sniffer trace uncertainty in Section~\ref{sec:uncertainty}.
 \section{Protocol Analyzer}
\label{sec:policy}

The \analyzer{} takes the protocol model which consists of a set of
communicating state machines, and outputs the packet success/loss sequences to
reach each system state. We first propose a approach to model wireless protocols
(\S~\ref{subsec:encoding}) and the algorithm to infer the packet success/loss
sequences (\S~\ref{subsec:reach}). We then describe our implementation on
\uppaal{} model checker (\S~\ref{subsec:uppaal}).

\subsection{Modeling Wireless Protocols}
\label{subsec:encoding}

Communication in wireless medium is broadcast at physical layer in nature---each
transmitted packet can be heard by every devices within the vicinity of the
transmitter. The packet contains a destination field to help non-designating
devices drop such packets and only the indented receiver delivers the packet to
upper layer. Due to factors such as interference and fading, each packet has a
non-zero loss probability.

\begin{figure}[t!]
  \centering
  \includegraphics[width=\figwidth{}]{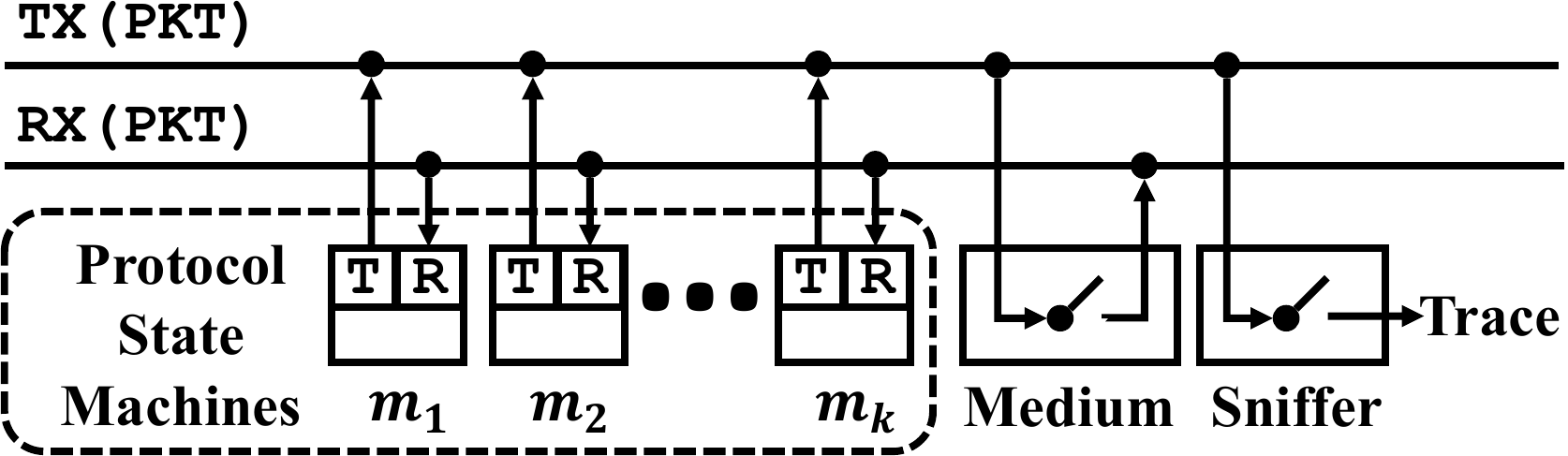}
  \caption{\textbf{Model for Wireless Protocols.} \tx{} and \rx{}
  represent common transmitting and receiving modules that are reused at each
state machine.}
  \label{fig:model}
\end{figure}

Based on these observations, we propose a {\it generic model} for wireless protocols,
as shown in Figure~\ref{fig:model}. Transmitting and receiving packets are
simulated by sending to and receiving from two shared \textit{synchronous} signal buses:
\texttt{TX(PKT)} and \texttt{RX(PKT)}. The wireless medium is modeled as a
\textit{switch} that controls the success or failure of each transmission at per-packet
basis. The wireless protocol to be modeled consists of a set of state machines,
$\m{1}, \m{2},\ldots,\m{k}$, that inter-operate with each other via the two
communication buses. The protocol state is a tuple of all the states at each
state machine, i.e., $\langle \m{1}.state, \m{2}.state,\ldots,
\m{k}.state\rangle$.

Assuming that the protocol itself is deterministic, the only non-determinism in the
model lies in the switch inside the \texttt{Medium} module. When querying certain
properties of the protocol model, such as state reachability,
the model checker tool can only manipulate this medium switch
for each packet when performing validation. The \texttt{Sniffer} module is
discussed in Section~\ref{sec:uncertainty}.

Note that there are two common functionalities that are used at each end
devices: transmitting and receiving a packet. We use the \tx{} and \rx{}
sub-modules to provide such abstractions. The packet transmission sub-module
(\tx{}) involves managing sequence number (for de-duplication at the receiver),
waiting for acknowledgment and performing retransmission when needed.  The
packet reception sub-module (\rx{}) is responsible for sending acknowledgment
packet and packet de-duplication. These two sub-modules, together with the
\texttt{Medium} and \texttt{Sniffer} models, are protocol independent and can be
provided as part of the verification framework.

The ability of modeling broadcast packets (such as beacons in 802.11) is limited
in our current formulation. Since there is only one medium switch for all
receivers, the broadcast packet can only be either received or missed by
\textit{all} devices. While in reality, the broadcast packet may be received by
a subset of devices but missed by others. However, this is not a fundamental
limitation, and can be mitigated by extending our formulation and adding a
medium switch to each participating device. Without loss of generality, we focus
on protocols that only contain single-cast packets in this paper.

\subsection{Jamming Policy Generation}
\label{subsec:reach}

\begin{algorithm}[t!]
    \caption{Jamming Policy Generator.}
    \label{alg:policy}
    \begin{algorithmic}[1]
        \Require protocol model $M$, state to reach $s$.
        \Ensure jamming policy to reach the state $s$, or \nil{} if
        the state is not reachable.
        \Procedure{JammingPolicy}{$M$, $s$}
        \State errorState $\gets s$ \label{alg:policy:neg}
        \State transitions $\gets$ \Call{ModelChecker}{$M$, errorState}
        \If{transitions = \nil{}}
            \State \Return \nil{}\Comment{state $s$ is not reachable}
        \EndIf
        \State policy $\gets$ []
        \For{t \textit{in} transitions}
            \State policy.append($\langle$t.pkt, t.medium\_switch$\rangle$)\label{alg:policy:trans}
        \EndFor
        \State \Return policy
        \EndProcedure
    \end{algorithmic}
\end{algorithm}

The output of \analyzer{} is the packet success/loss sequences that drive the
protocol model into each state. We call such sequences \textit{jamming policy},
as they provide instructions of which packet to jam or pass for \jammer{} that
we will describe in Section~\ref{sec:jammer}.

With our proposed modeling approach, the key idea of inferring jamming policy to
reach a target state is to reduce the problem to the well-known model checking
problem of determining if an error state $s$ is reachable in a model
$M$~\cite{model-checking-book}.

Algorithm~\ref{alg:policy} shows the jamming policy generator algorithm.
In order to infer the jamming policy to reach the target state, we instruct the
model checker to verify the property that claims the target state is not
reachable (L\ref{alg:policy:neg}). If the state is reachable, model checker will
come up with a counter example which consists of the state transitions from the
initial state to the target state. By examining the packet and the state of
medium switch associated with the transition (L\ref{alg:policy:trans}), we can
infer the jamming policy that drives the system to target state. We then apply
Algorithm~\ref{alg:policy} for each system state, and obtain corresponding
jamming policies for each reachable system state.

\subsection{Implementation on \uppaal{}}
\label{subsec:uppaal}

The proposed modeling methodology of wireless protocols and algorithm for
inferring jamming policies are generic, and can be realized in any model
checking tool that provides synchronization capability between interacting
models. Next, we describe one implementation on \uppaal{} model checker.

\subsubsection{\uppaal{} Primer}

\uppaal{}~\cite{bengtsson1996uppaal,behrmann2006uppaal} is a model checking suite for
verifying real-time systems modeled as networks of timed
automata~\cite{alur1994theory}.  The suite contains the \uppaal{} language
specification and a model checker implemented using constraint-solving
techniques.

\begin{figure}[t!]
    \centering
    \includegraphics[width=\figwidth{}]{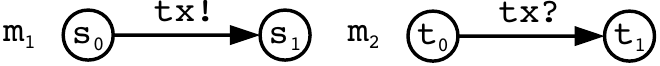}
    \caption{\textbf{Channel Synchronization in \uppaal{}. }}
    \label{fig:uppaal:chan}
\end{figure}

\uppaal{} provides shared variables and synchronization capability via
{\it channel} variables, making it easy to model packet
exchange. An example is shown in Figure~\ref{fig:uppaal:chan}. A channel
variable \texttt{tx} is defined to synchronize the transitions in two state
machines, $\m{1}$ and $\m{2}$. In particular, the $s_0 \rightarrow s_1$
transition labeled as \texttt{tx!} in $\m{1}$ synchronizes with the $t_0
\rightarrow t_1$ transition labeled as \texttt{tx?} in $\m{2}$. Therefore, a
packet can be ``transmitted'' from $\m{1}$ to $\m{2}$ by writing a shared
variable in the $s_0\rightarrow s_1$ transition in $\m{1}$, and reading the same
shared variable in the $t_0\rightarrow t_1$ transition in $\m{2}$.

\begin{lstlisting}[
  float=t,
  language={[GUI]Uppaal},   columns={[l]flexible},
  basewidth=.5em,
  frameround=fftt, frame=shadowbox, rulesepcolor=\color{gray},
  caption={\textbf{State Transition in \uppaal{}.}},
  belowskip=0mm,
  label=lst:transition
]
// transition from s_i to s_j
s_i -> s_j {
  select i : int [0, 1];     // Selection
  guard c < 5;                   // Guard
  sync signal!;        // Synchronization
  assign c = 0;}                // Update
\end{lstlisting}

Listing~\ref{lst:transition} shows the 4 optional components of a state transition in
\uppaal{}. The \textit{Selection} statement non-deterministically binds a local
identifier to a value in a given range. The state transition is enabled if and
only if the \textit{Guard} statement is evaluated to true. Transitions labeled
as \textit{Synchronization} pairs synchronize with each other. For instance, a
transition labeled as \texttt{signal!} triggers another transition
labeled as \texttt{signal?}. The \textit{Update} statements are
evaluated when the transition happens. The side-effect of the \textit{Update}
statements update the state of the model.

\subsubsection{Modeling Using \uppaal{}}

Next, we show how the \tx{}, \rx{} and \texttt{Medium} models in our
formulation can be implemented in \uppaal{}.

\begin{lstlisting}[
  float=t!,
  language={[GUI]Uppaal},   columns={[l]flexible},
  basewidth=.5em,
  frameround=fftt, frame=shadowbox, rulesepcolor=\color{gray},
  caption={\textbf{Modeling \texttt{T}, \texttt{R} and \texttt{Medium} using \uppaal{}.}
  The body of \texttt{T} and \texttt{R} is omitted for sake of space.},
  belowskip=0mm,
  label=lst:dot11
]
const int NUM_DEVICE = 2;

// interface of the T module
chan mac_tx_start[NUM_DEVICE]
pkt_t pkt_to_send[NUM_DEVICE];
chan mac_tx_done[NUM_DEVICE];
bool tx_ok[NUM_DEVICE];

process T(int id) {
  // listen for mac_tx_start[id]
  // and transmit pkt_to_send[id]
  ...  }

// interface of the R module
chan mac_rx_end[NUM_DEVICE];
pkt_t pkt_recvd[NUM_DEVICE];

process R(int id) {
  // listen for the phy_rx_end
  // and trigger mac_rx_end[id] on
  // packet received
  ...  }

// interface of the Medium module
chan phy_tx_start;
chan phy_rx_end;
pkt_t pkt_in_air;

process Medium() {
  bool loss;
  state
    s_init, s_got_pkt;
  transitions
    s_init -> s_got_pkt @\label{lst:dot11:m:select}@
    { select i: int [0, 1];
      sync phy_tx_start?;
      assign loss = i == 0; },
    s_got_pkt -> s_init
    { guard loss; },
    s_got_pkt -> s_init @\label{lst:dot11:m:succ}@
    { guard !loss;
      sync phy_rx_end!; }; }
\end{lstlisting}

Listing~\ref{lst:dot11} shows the \uppaal{} encoding of the \tx{}, \rx{} and
\texttt{Medium} model in our formulation. The interface to the \tx{} model is
two synchronization channels for each device (\texttt{mac\_tx\_start} and
\texttt{mac\_tx\_done}), together with two shared variables
(\texttt{pkt\_to\_send} and \texttt{tx\_ok}) for passing data. Internally the
\texttt{T} model handles sequence number assignment, expecting acknowledgments
and retransmissions if needed.

Similarly, the interface of the \rx{} model is one channel
(\texttt{mac\_rx\_end}) and one shared variable (\texttt{pkt\_recvd}) for each
device. The \texttt{R} model handles sending acknowledgment and packet
de-duplication. Both the \tx{} and \rx{} model are parameterized with an integer
identifier so that they can be reused by multiple device models.

Inside the \texttt{Medium} model, for each transmitted packet, it
non-deterministically decides whether the packet is lost
(\lstline{lst:dot11:m:select}), and only sends the \texttt{phy\_rx\_end} signal if
the packet not lost (\lstline{lst:dot11:m:succ}). Therefore, the
\texttt{Medium} model acts as a switch between the \texttt{phy\_tx\_start} and
the \texttt{phy\_rx\_end} signals.

\begin{lstlisting}[
  float=t!,
  language={[GUI]Uppaal},   columns={[l]flexible},
  basewidth=.5em,
  frameround=fftt, frame=shadowbox, rulesepcolor=\color{gray},
  caption={\textbf{Example Modeling of Client Authentication State Machine Using
  \uppaal{}.}},
  belowskip=0mm,
  label=lst:auth
]
s_0 -> s_1 { @\label{lst:auth:c:req}@
  sync mac_tx_start[client_id]!;
  assign pkt_to_send[client_id].type = AUTH_REQ;@\label{lst:auth:pkt}@
    pkt_to_send[id].dest = ap_id; },
\end{lstlisting}

With the help of the \tx{} and \rx{} models, it is easy to encode the up layer
state machines in \uppaal{} language. For instance, the encoding of
$s_0\rightarrow s_1$ transition in the client state machine is shown in
Listing~\ref{lst:auth}. To send the \authreq{} packet, the client model only
needs to set the packet to be sent (L\ref{lst:auth:pkt}), then signals the
corresponding \texttt{max\_tx\_start} signal. The rest of sending logic is
handled by the \texttt{T} module.

\subsubsection{Jamming Policy Generator}

In \uppaal{}, the property to be verified is expressed using the
TCTL~\cite{alur1990model} query language. In particular, given a property $\phi$,
the query \texttt{A[]$\phi$} asks whether $\phi$ holds for all possible
execution paths. This query can be used to generate the jamming policies.

\sloppy{  For instance, the \texttt{A[] !(Client.s\_4 \&\& AP.t\_4)} query
  asserts that the system state \texttt{$\langle$Client.s\_4,
  AP.s\_4$\rangle$} is not reachable. There are two possible outcomes from
  model checking such a property: either the answer is \texttt{yes} and the
  state is indeed not reachable, or the answer is \texttt{no} and \uppaal{} will
  also provide one transition sequence from the initial state to the target
  state. By examining the value of the \texttt{pkt\_in\_air} and the
  \texttt{Medium.loss} variables at each intermediate state, we can extract the
  jamming policy in order to reach the target state.
}

We have implemented \analyzer{} for \uppaal{} using Python. It first parses the
\uppaal{} model to extract the models and their states. For every combination of
the model states, it generates the un-reachability query described above and runs the \uppaal{}
model checker.  If the query is not satisfied (i.e., the state is reachable), it
parses the \uppaal{} output and extracts the packet success/failure sequence. The
output of the \analyzer{} is a list of reachable system states and the packet
sequence in order to drive the system to the target state.
 \section{PacketSniper}
\label{sec:jammer}

With the jamming policy to reach each protocol state, we now
describe the design and two prototype implementations of \jammer{} to execute
the jamming policy.

\subsection{Design Requirements}

\begin{figure}[t!]
  \centering
  \includegraphics[width=\figwidth{}]{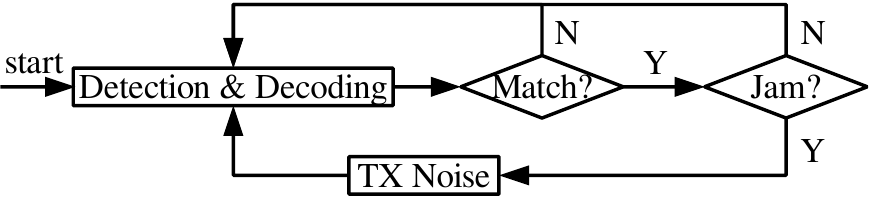}
  \caption{\textbf{Work Flow of \jammer{}.}}
  \label{fig:jammer}
  \vspace*{0mm}
\end{figure}

To ensure the integrity of the transmitted packets, wireless protocols usually
include various forms of checksums in the packet to help the receiver to detect
bit errors. In wireless protocols such \wifi{}, packets with checksum errors are
discarded by the receiver. The idea of \jammer{} is to detect the beginning of
packet transmission in real time, and disrupt the transmission just in time
before it ends to cause the checksum to fail at the receiver side.
Figure~\ref{fig:jammer} shows the overall work flow of \jammer{}. The \jammer{}
continuously monitors the wireless medium, and starts decoding packets upon
detection. It then performs matching of the packet with predefined jamming
policies and sends jamming signals if the packet is labeled as failure in the
policy.

To achieve this goal, the \jammer{} needs to first decode certain packet
attributes, such as source, destination and packet type.
Therefore, previous physical layer jamming
methods~\cite{clancy2011efficient,park2003effect,la2012jamming} are not
applicable as they can only blindly jam the wireless medium.
The \jammer{} then needs to transmit the jamming signal before the original
packet transmission ends in order to disrupt the transmission. This poses very
strict timing requirements as the duration of packet transmissions in wireless
communication is often in the order of microseconds.

Next, we first describe a set of design requirements and challenges of
\jammer{}. We then describe two prototype \jammer{} implementations for 802.11
protocol on Universal Software Radio Peripheral (USRP) N210 and SORA platform.

\subsubsection{Jamming Latency}
\label{subsubsec:latency}

The \jammer{} must transmit the jamming signal before the end of packet
transmission to jam the packet at the receiver side.
This requires the \jammer{} to perform the packet detection, decoding and
decision making process within a portion of the transmission duration. There
are several MAC protocol implementations on Software Define Radio (SDR)
platforms~\cite{nychis2009enabling,schmid2007experimental}. However, the round
trip latency between the SDR frontend and the host PC is usually in order of
hundreds of milliseconds, which are prohibitively longer than the requirements.
It is clear that the core logic of \jammer{} must be placed as close to the
radio frontend as possible to meet the tight timing requirements.

In addition, certain management or action packets, such as the acknowledgment
(ACK), Request to Send (RTS) and Clear to Send (CTS) packets in 802.11 protocol,
only contains very minimum protocol attributes and no data payloads, making them
extremely difficult to jam. The \jammer{} needs to take advantage of
certain protocol semantics in order to jam such short packets.

\subsubsection{Protocol Attributes}
\label{subsubsec:tradeoff}

\jammer{} needs to decode enough protocol attributes of the packet to help make
the jamming decision, yet must stop decoding as early as possible so that the
jamming signal can overlap enough with the original transmission. Therefore, the
number of protocol attributes to decode represents a trade-off between jamming
policy granularity and jamming probability. The more attributes to decode, the
more fine granularity the jamming policy can express, yet the probability of
successfully jam the packet is lower.

\subsubsection{Decoding Errors}

As wireless receivers, the \jammer{} could also have decoding
errors.  This may cause the \jammer{} to miss the packets to be jammed, or jam
the wrong packets. While the \jammer{} should try its best to execute the
jamming policy, such jamming misbehaviors create uncertainty in later
verification phase.

\subsection{Implementation on USRP}

We now describe our prototype implementation of \jammer{} for 802.11 protocol on
USRP N210 SDR platform. Because of the tight timing requirements, we choose to
implement the core jamming logic in the on-board FPGA to eliminate the long
round-trip latency suffered by most SDR implementations.

\begin{figure}[t!]
  \centering
  \includegraphics[width=\figwidth{}]{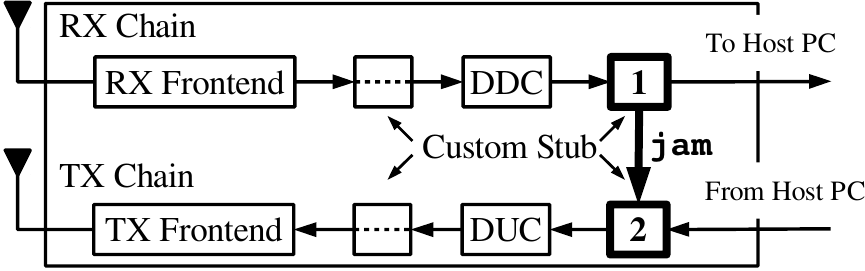}
  \caption{\textbf{USRP N210 FPGA Diagram.} Our implementation is in the two
  bold blocks marked as 1 and 2.  DDC/DUC stands for Digital Down/Up Converter.
}
  \label{fig:fpga}
\end{figure}

The USRP N210 platform contains an on-board FPGA for time-sensitive
baseband tasks, such as I/Q balance, digital down/up converting. The current
resource utilization of the N210 FPGA is only around 25\%, leaving enough room
for implementing custom logic on board.

Figure~\ref{fig:fpga} shows a simplified diagram of the
USRP N210 FPGA. There are two RF chains: one for receiving the RF signals and
sending them to host PC (the RX Chain), and the other for transmitting the RF
signals from host PC to the air (the TX Chain). On both RF chains, there are
stubs reserved for implementing custom logic. By default, these stubs are only
pass-through, thus have no effect on the signal processing.

We have modified the custom stubs to implement \jammer{}, as shown in the
two bold boxes labeled as 1 and 2. In stub 1, we implemented a full 802.11a/g/n
OFDM receiver. The tasks include: short preamble detection~\cite{liu2003design}
and coarse frequency offset correction~\cite{sourour2004frequency}, long
preamble detection for symbol alignment and fine frequency offset
correction~\cite{sourour2004frequency}, OFDM decoding consists of demodulation,
deinterleaving, Viterbi decoding and descrambling.

In addition, there is a jamming filter module in stub 1 that compares the
decoded packets with pre-configured jamming policies to determine the jamming
decision. The output of stub 1 is a binary signal, called \texttt{jam}. When the
\texttt{jam} signal is asserted, stub 2 starts transmitting pre-configured
jamming signals through the TX chain.

The overall implementation includes 4320 lines of Verilog code for FPGA
implementation, and 2265 lines of Python code for generating look up tables and
cross validation of the FPGA implementation.

\subsection{Implementation On SORA}

Another promising SDR platform for \jammer{} is
SORA~\cite{tan2011sora}. Currently, SORA provides a full implementation of 802.11a/b/g and
part of 802.11n up to 2 spatial streams. SORA utilizes real-time threads to meet
the strict timing requirements of wireless protocols on general purpose
operating system (Windows).

We build our \jammer{} prototype on top of the packet decoder in SORA. More
specifically, we instruct the packet decoder to only decode first few bytes of
the packet, which are then used to perform jamming filter matching. Once a
positive jamming decision is made, we notify the SORA firmware to transmit
pre-loaded jamming signal.

Compared to the FPGA implementation on USRP, SORA introduces additional jamming
latency due to the data exchange between SORA firmware and host CPU, yet can be
potentially used to support latest wireless standards (e.g., 802.11ac) and MIMO
operations.

\subsection{Jamming Policy Format}
\label{subsubsec:policy}

Jamming policy represents the interface between \analyzer{}
and \jammer{}, and contains two parts: filter and action.  The jamming filter
defines \textit{what} packets to jam, while the action defines \textit{when} to
jam the packet. For each packet success/failure sequence output by
\analyzer{}, we first convert it into a mapping from packet filter
to a list of actions. Next, we describe the format of the
filter and action respectively.

We note that wireless protocol attributes typically include certain physical
layer properties, such as packet length and encoding rate, and the packet header
that contains source and destination information, type, control flags and so on.
Therefore, we restrict the jamming filters to these three attributes.

\begin{lstlisting}[
  float=t,
  language={[GUI]Uppaal},   columns={[l]flexible},
  numbers=none,
  basewidth=.5em,
    caption={\textbf{BNF Definition of Jamming Filter.}},
  belowskip=0mm,
  label=lst:filter
]
Filter ::= "true" | Predicate | Filter && Filter
Predicate ::= x <= const | x >= const
x ::= "length" | "rate" | "header"
\end{lstlisting}

Listing~\ref{lst:filter} shows the BNF specification of our proposed filter
syntax. A jamming filter is a conjunction of predicates, each of which performs
filtering on one of the three attributes. The \texttt{length} and \texttt{rate}
attribute are easy to understand. There are two points worth noticing about the
\texttt{header} attribute. First, as explained earlier in
Section~\ref{subsubsec:tradeoff}, the length of the header represents the
trade-off between filter granularity and jamming latency. Here we do not use a
fixed length but instead make the length as an configuration parameter at run
time.  Second, since the \texttt{header} is a continuous chunk of bytes while
the actual protocol attributes may scatter in different segments of the headers,
a special notation of \textit{don't care} byte is used to skip the
non-interested bytes.

The jamming policy also contains a list of actions for the corresponding jamming filter. The
possible actions are: \texttt{Jam}, \texttt{Pass}, and \texttt{JamNext}. The
\texttt{JamNext} action instructs the \jammer{} to skip the current packet but
jam the next detected packet. This action is introduced to overcome the
challenge of jamming short packets as described in
Section~\ref{subsubsec:latency}. Our observation is that such short packets
(such as acknowledgment or CTS packets) are typically the response of certain
other packets. Although the short packets themselves are difficult to jam, it is
easy for the \jammer{} to predict their transmissions beforehand and sends the
jamming signal as soon as detecting the next packet transmission, without
needing to decode its content. We acknowledge that this action may not cover
all cases for short packets, such as the RTS packet in 802.11.

 \section{Sniffer Trace Verification}
\label{sec:uncertainty}

The final step is to validate whether the DUT's behavior is consistent with the
protocol model. We first explain the need of using sniffer as vantage
point for verification (\S~\ref{subsec:vantage}), then discuss the challenges of
sniffer trace verification and our proposed modeling methods
(\S~\ref{subsec:sniffer}). Finally we show how our modeling can be implemented
in \uppaal{} (\S~\ref{subsec:uppaal_sniffer}).

\subsection{Sniffer as Vantage Point}
\label{subsec:vantage}

The broadcast nature of wireless medium makes it possible to observe the
DUT's packet exchanges from external devices, or wireless sniffers.
Using wireless sniffer as vantage point is more of a requirement of practical
constraints rather than a design choice. Due to fine-grained timing requirement in
wireless protocols, the implementations are often placed as close as to the
hardware in the form of firmwares. Furthermore, these firmwares are usually
proprietary, making it difficult to instrument the implementation to collect
packet exchange traces or event logs. Even when such instrumentation capability is
available, the resource constraints in most embedded or IoT devices make it
infeasible to directly collect packet traces from the devices under test.

As shown in Figure~\ref{fig:model}, the sniffer can be modeled as a passive
observer (\texttt{Sniffer}\footnote{We use ``sniffer'' to refer to the
physical wireless sniffer devices, and use ``\texttt{Sniffer}`` to refer to the
model in Figure~\ref{fig:model}.}) of the \texttt{TX(PKT)} signal. Just as
regular receivers, the sniffer could also miss packets, thus there is another
switch inside the \texttt{Sniffer} model. Note that the two switches inside the
\texttt{Medium} and the \texttt{Sniffer} models are \textit{independent} from
each other. For instance, a packet could be received by the designating receiver
but be missed by the sniffer, and vice versa. Also note that although we model
sniffer as one logical entity, in practice it may consist of multiple physical
sniffer devices and their time-synchronized
traces~\cite{Cheng:2006:JSP:1159913.1159920,bahl2006enhancing,Mahajan:2006:AMB:1159913.1159923}
together form the logical sniffer's observation in our model.

\subsection{Verifying Sniffer Trace}
\label{subsec:sniffer}

Given a sniffer trace that represents the \textit{implementation's} behavior,
the verification problem is to check whether it is one of the legal
observations of the \texttt{Sniffer} model. If the
answer is no, then a violation of the protocol can be claimed. On the other
hand, however, a complete protocol compliance can not be declared even if the
answer of the verification problem is yes. This is because a positive answer
only means the implementation \textit{could have} behaved according to the
protocol, but it is also possible that the implementation violated the protocol
but the violating behavior was missed by the sniffer. This is known as the
\textit{sniffer uncertainty} problem first described by \rv{}~\cite{shi-rv16}.

\begin{figure}[t!]
  \centering
  \includegraphics[width=\figwidth{}]{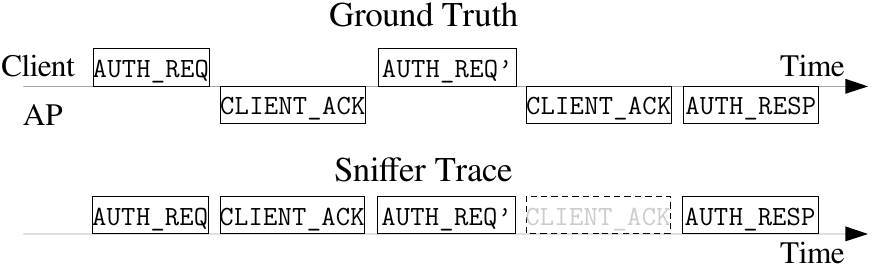}
  \caption{\textbf{Example of Sniffer Trace Uncertainty.}}
  \label{fig:uncertainty}
\end{figure}

More specifically, the sniffer may miss packets that represent protocol
violations. Also, the sniffer can not determine whether an observed packet was
actually received by the receiver. Figure~\ref{fig:uncertainty} shows an example
of such uncertainty. Note that the first \clientack{} packet was missed by the
client but was received by the sniffer. A na\"ive validation of this sniffer
trace would report a violation since the client retransmit the \authreq{} packet
after \textit{receiving} the \clientack{} packet.

Using sniffer traces for verification, we share the same limitations
with~\cite{shi-rv16} caused by sniffer uncertainty. However, in our formulation,
the sniffer uncertainty is implicitly explored by the model checker in toggling
the two switches inside the \texttt{Medium} and \texttt{Sniffer} models, whereas
it has be to explicitly expressed using the augmented transitions
in~\cite{shi-rv16}.  More specifically, a $\langle\text{\texttt{true,
false}}\rangle$ value of the $\langle\text{\texttt{Medium.loss,
Sniffer.loss}}\rangle$ tuple represents the case when a packet is received by
the receiver but missed by the sniffer.  Similarly, a
$\langle\text{\texttt{false, true}}\rangle$ value corresponds to the case when a
packet is missed by the receiver but received by the sniffer.

In verifying sniffer traces in our framework, the model checker will try all 4
values of the loss variable tuple for each transmitted packet in the attempt to
accept the sniffer trace. We argue that by considering sniffer as part of
the model in our formulation, the sniffer uncertainty can be explored in a
systematic manner by an off-the-shelf model checker rather than a specialized checker~\cite{shi-rv16}.

We also note that the number of sniffer losses that the model checker is allowed
to use must be bounded. Otherwise the model checker will either indefinitely
infer packets missed by the sniffer without making progress on the actual
sniffer trace, or yield artificial traces that contains too many sniffer losses
to be practically meaningful. As observed in~\cite{shi-rv16}, the number of
sniffer losses required for the model checker to accept the trace, denoted as
$k$, is inversely proportional with the confidence of the trace correctness.
Intuitively, the larger the $k$, the more missing packets the model check needs to
\textit{guess} to accept the trace, thus  less confidence that the trace
represents correct protocol behavior.

\subsection{Modeling Sniffer in \uppaal{}}
\label{subsec:uppaal_sniffer}

Next, we show how the \texttt{Sniffer} model and the sniffer trace verification
can be done in \uppaal{}. The key idea is to embed the sniffer
trace as a constant array, and assert that the \texttt{Sniffer} model must observe
the packets sequentially.

\begin{lstlisting}[
  float=t,
  language={[GUI]Uppaal},   columns={[l]flexible},
  basewidth=.5em,
  frameround=fftt, frame=shadowbox, rulesepcolor=\color{gray},
  caption={\textbf{Modeling \texttt{Sniffer} using \uppaal{}.}},
  label=lst:sniffer
]
// sniffer trace to be verified
const int LEN = 10;
const pkt_t TRACE[LEN] = {...};

process Sniffer() {
  bool loss;
  int idx = 0; @\label{lst:sniffer:idx}@
  state
    s_init, s_got_pkt;
  transitions
    s_init -> s_got_pkt { @\label{lst:sniffer:tap}@
      select i: int [0, 1];
      sync phy_tx_start?;
      assign loss = i == 0; },
    s_got_pkt -> s_init {
      guard loss; },
    s_got_pkt -> s_init { @\label{lst:sniffer:eq}@
      guard !loss && idx < LEN && @\label{lst:sniffer:guard}@
        pkt_in_air == TRACE[idx];
      assign idx++; }; }
\end{lstlisting}

Listing~\ref{lst:sniffer} shows the \uppaal{} encoding of the \texttt{Sniffer}
model. A constant array of packets, \texttt{TRACE}, is defined. The elements and
length of the array is populated from the sniffer trace to be verified. The
\texttt{Sniffer} model taps in the \texttt{phy\_tx\_start} signal
(\lstline{lst:sniffer:tap}) to monitor every packet transmission, and
non-deterministically decides whether the transmitted packet was received by the
sniffer or not. The \texttt{Sniffer} model also maintains an index to the
\texttt{TRACE} array (\lstline{lst:sniffer:idx}). If the transmitted packet was
received by the sniffer, then it must be the next expected packet in the sniffer
trace (\lstline{lst:sniffer:eq}). This way, we are asserting that the
\texttt{Sniffer} model should observe the same packet trace as the physical
sniffer. The bound on number of sniffer loss, which is omitted for sake of
space, can be placed in the guard of the
\texttt{s\_got\_pkt}$\rightarrow$\texttt{s\_init} transition
(\lstline{lst:sniffer:guard}).

The verification is then performed by the \texttt{E<> Sniffer.idx == LEN} query,
which states that there should exist at least one
transition path along which the \texttt{idx} value eventually reaches
\texttt{LEN}. If the query is satisfied, then the sniffer trace is indeed one of
the possible traces observed by the \texttt{Sniffer} model.
 \section{Evaluation}
\label{sec:eval}

In this section, we perform both micro benchmarks on \jammer{} as well as
end-to-end evaluation using the 802.11 link setup protocol.

\subsection{Micro Benchmarks of \jammer{}}

\subsubsection{Jamming Latency on USRP}

As described in Section~\ref{subsubsec:latency}, one major challenge of
realizing \jammer{} is the critical timing requirement: the decoding and filter
matching process must be completed before the end of packet transmission. To
quantify this requirement, we define the \textit{Jamming Latency} as the duration
between the beginning of the packet and the time when the jamming signal
is transmitted. The latency must be shorter than the duration of the packet in
order for the jamming signal to overlap with the packet. Next, we evaluate the
jamming latency on our USRP N210 prototype implementation of \jammer{}.

\begin{figure}[t!]
  \centering
  \includegraphics[width=\figwidth{}]{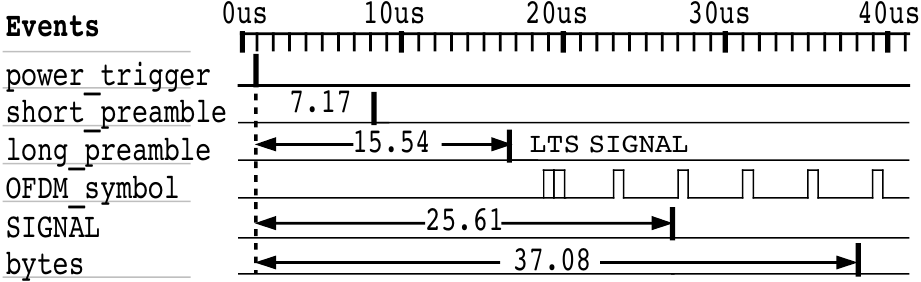}
  \caption{\textbf{Timeline of Various Decoding Stages of USRP N210 \jammer{}
  Implementation.}}
  \label{fig:wave}
\end{figure}

Figure~\ref{fig:wave} shows the timeline of various stages of the decoding a
6~Mbps 802.11a OFDM packet on the USRP N210. The \texttt{power\_trigger} event
is triggered when the receiving power level significantly increases, and marks
the beginning of packet transmission. The \texttt{short\_preamble} and
\texttt{long\_preamble} events are triggered when the presence of
short and long preamble are detected respectively. The latency of these two
events (7.17~\us{} and 15.54~\us{}) corresponds to the fact that the short and
long preambles are 8~\us{} each in duration. For jamming policies that are only
interested in the presence of a packet (such as the \texttt{JamNext} action), a
jamming decision can be made at this stage, and all following decoding steps can
be skipped.

The \texttt{SIGNAL} field of the packet, which contains physical layer
properties such as packet length and encoding rate, is available after
25.61~\us{}. For jamming filters that only contain these two attributes, their
matching results can be concluded, and their jamming decision can be made at
this point.  No further decoding is required.

The first byte of the packet is available after 37.08~\us{} from the beginning
of the packet transmission. Depending on the header length in the jamming
filter, more bytes might be needed to complete the filter matching process. The
incurred extra latency depends on the exact encoding rate of the packet. The higher
encoding rate, the more data bits each OFDM symbol contains, thus the shorter
decoding latency.

To summarize, the jamming latency in our \jammer{} implementation can be as
short as 8~\us{} for certain jamming actions, and as long as 37~\us{} plus
additional decoding time for jamming filters that contain header bytes.

Once the jamming decision is made, the \texttt{jam} signal is asserted and
jamming signals are transmitted through the TX chain. The RX/TX turn-around time
on USRP N210 is negligible (in order of nanoseconds) as the RX and TX chain are
independent from each other.

\begin{figure}[t!]
  \includegraphics[width=\figwidth{}]{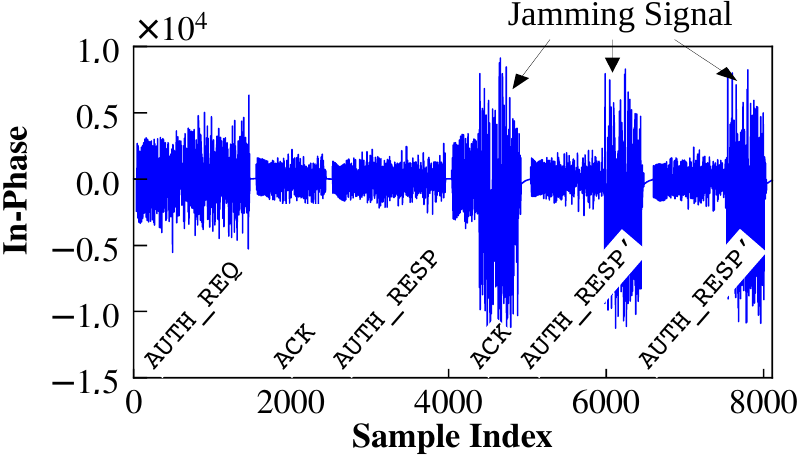}
  \caption{\textbf{Jamming in Action.} The first jam signal was sent
  using the \texttt{JamNext} action, while the following two were using the
\texttt{Jam} action.}
\label{fig:jam}
\end{figure}

Figure~\ref{fig:jam} shows the in-phase signals captured during one jamming
session we performed for the 802.11 link setup protocol. The particular jamming
policy was to jam the acknowledgment packet of the \authresp{} packet and all
following retransmissions of the \authresp{} packet. Since the acknowledgment
packets are short, we utilize the \texttt{JamNext} action to jam the
acknowledgment of the \authresp{} packet. As can be clearly seen in
Figure~\ref{fig:jam}, the jamming signals, whose magnitudes are significantly
larger than the original signals, are transmitted in the middle of packet
transmissions. In addition, the jamming latency for the acknowledgment packet is
short than the two following \authresp{}' packets. This is because the jamming
decision was made immediately after the \texttt{short\_preamble\_detected} event
for the acknowledgment packet. While for the two \authresp{}' packets, the jamming
decision was made after 10 header bytes were decoded.

\subsubsection{Jamming Latency on SORA}

\begin{figure}[t!]
  \centering
  \includegraphics[width=\figwidth{}]{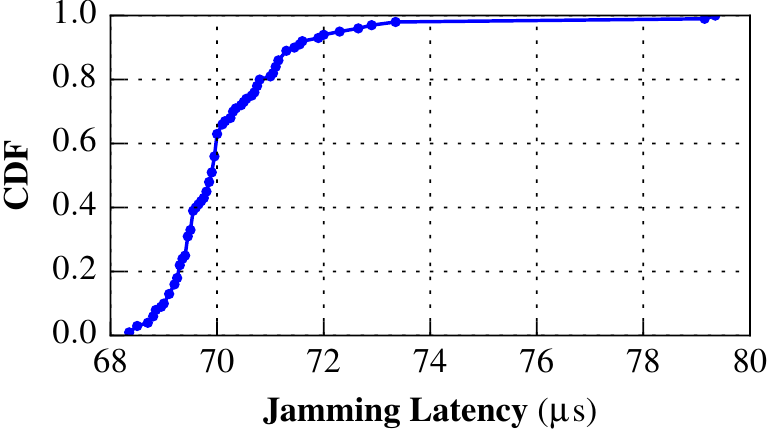}
  \caption{\textbf{CDF of Jamming Latency on SORA.}}
  \label{fig:sora_delay}
\end{figure}

To measure jamming latency on SORA, we leverage another radio hardware to
capture the jammed packets over the air. Since jamming signal is LTS, we perform
cross-correlation with LTS to detect both the start of a jammed packet and the
jamming signal. In our experiment, the data packets are transmitted at 6Mbps.

Figure~\ref{fig:sora_delay} shows CDF of the jamming latency on SORA for
100 jammed packets. The medium latency is around 70~\us{}, and is
stable within the 70$\pm$2~\us{} range. Occasionally, jamming latency may jump
to 80\us{}, because sometimes SORA thread needs to release the CPU core to
prevent from blocking hardware interrupts.

\subsubsection{Jamming Capability}

Finally, we evaluate \jammer{}'s capability of jamming packets. We focus on the
USRP implementation since our SORA based prototype is still work in progress

We set up a transmitter-receiver pair in a conducted environment, and configure
the transmitter to send a constant number (100) of data packets using the
\texttt{ping} command. The transmitter data rate is fixed at 6~Mbps.  The jammer
is configured to jam all the non-retransmission packets from the transmitter.
There is no path-loss between the jammer and the receiver. We then use three
pass-loss values between the transmitter and receiver: 0~dB, 20~dB, 40~dB. In
these settings, the RSSIs at the receiver side are -20~dBm, -40~dBm, -60~dBm
respectively. We then vary the transmission power of the USRP N210 from -10~dBm
to 20~dBm with 1~dBm step.

\begin{figure}[t!]
  \centering
  \includegraphics[width=\figwidth{}]{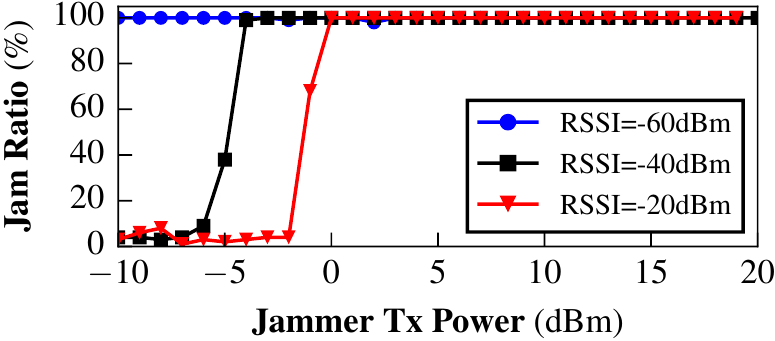}
  \caption{\textbf{Jamming Ratio of \jammer{} Implementation on USRP
  N210.}}
  \label{fig:ratio}
\end{figure}

Figure~\ref{fig:ratio} shows the jamming success ratio under the three path loss
settings. As expected, when the RSSI at the receiver side is low, the jammer can
easily jam all packets even at lowest transmission power. As the RSSI increases,
the required jammer transmission power increases accordingly. After certain
cut-off point, the jammer was able to jam nearly all the packets.

 \subsection{End-to-End Evaluation}

We perform an end-to-end evaluation of our framework through a case study of the
802.11 link setup protocol. The results show that our framework can increase
state coverage by 2X (\S~\ref{subsubsec:coverage}) and find violations that
otherwise do not manifest (\S~\ref{subsubsec:vio}).

\subsubsection{Protocol Modeling}

We first give an overview of our modeling of the 802.11 link setup protocol
using UPPAAL. As informally described in Section~\ref{subsec:protocol}, the
protocol consists of three stages: authentication, association and 802.11X
authentication. Listing~\ref{lst:auth} shows our modeling of the first
authentication stage. We have modeled all three stages using UPPAAL. The
\texttt{Client} model consists of 13 states, and the \texttt{AP} model consists
of 17 states. The full model consists of 544 lines of UPPAAL code, 275 of which are for
the generic \texttt{T} and \texttt{R} modeling.

We then feed the model to \analyzer{}. Among the all possible 221 ($13\times
17$) system states, 73 states are reachable, and 40 reachable states involves
packet failures in their state transition sequence.

\subsubsection{Experiment Setup}

We use conducted setup for our end-to-end evaluation to eliminate external
interferences. The client and the AP are put
inside RF shield boxes, and their only way of communication is via a coaxial
cable that connects their antenna ports. We use two TP-LINK WDR3500 routers as
testing devices since they have detachable antennas. One router is configured in
client mode, and the other is configured in AP mode. Both routers runs
OpenWRT 15.05 firmware so that we can remotely control the link setup
process. The source code of OpenWRT is available, enabling us to examine any
potential implementation bugs we found later.

For each of the 73 jamming policies, we initiate
the link setup process on the client for 10 times, and run the \jammer{} with
corresponding policy during each link setup session. We utilize the packet log
of the \jammer{} as sniffer trace in later verification stage.

\subsubsection{State Reachability}
\label{subsubsec:coverage}

\begin{table}[t!]
  \centering
  \caption{\textbf{Summary of Sniffer Trace Verification Results.}}
  \label{tab:state}
  \begin{tabular}{||c||c|c|c||}
    \hline
    $k$ & Accepted & Rejected & Reached\\\hline\hline
    0 & 155 & 575 (79\%) & 289 \\\hline
    2 & 498 & 232 (32\%) & 592\\\hline
    4 & 617 & 113 (15\%) & 699 \\\hline
    6 & 641 & 89 (12\%) & 723 \\\hline
    8 & 663 & 67 (9\%)  & 729 \\\hline
    10 & 665 & 65 (9\%) & 729 \\\hline
  \end{tabular}
\end{table}

We perform verification on the resulting 730 ($73\times10$) traces using the
\texttt{E<> Sniffer.idx == LEN} query as explained in
Section~\ref{subsec:uppaal_sniffer}.  A trace is ``Accepted'' if the query is
satisfied, otherwise is ``Rejected'', which represents potential implementation
violations and needs to be further examined. In addition, we also verify if the
target system state was reached during the verification (``Reached'').

Table~\ref{tab:state} shows the verification results for various $k$ values.
Recall that $k$ is the maximum number of sniffer missed packets that the model
checker would tolerate before rejecting the trace. Overall, we can clearly
observe the trend that when $k$ increases, the model checker accepts more
sniffer traces, and more traces reached the target system state.

When $k=0$, which effectively disables the model checker's ability to infer
missing packets, 79\% of the traces were rejected. This shows the necessity of
tolerating sniffer losses in our formulation, as otherwise intensive labor effort is
required to examine the rejected traces to prune out false alarms causes by
sniffer missing packets.

When $k=10$, the number of rejected traces reduces to 65. We inspected
these traces and found that main cause was the device's internal queueing effect
that was not part of our model (this was a design decision to keep the model simple). For instance, the client may enqueue a new round
of authentication if the first \authreq{} packet was not acknowledged but indeed
received by the AP. This results an interleaving authentication packet sequence
from the same client, which will be rejected by the model checker. We also observe
several violations that are potentially caused by implementation misbehaviors,
which we report later in Section~\ref{subsubsec:vio}.

Additionally, when $k=10$, for 729 out of the 730 traces, the target system
state was reached in verifying the sniffer trace. The only failing trace was
caused by a violation (discussed in \S~\ref{subsubsec:vio}) in the association
stage while the system state is in the following 802.11X authentication stage.

Our framework enables testers to iteratively refine the verification process.
Starting with a high $k$ value, the tester can narrow down to a small subset of
traces that have high likelihood of violation. After either refining the original
model or identifying true protocol violations, the tester can then decrease $k$
and repeat the process.

Finally, we performed a set of baseline experiments without \jammer{}.  More
specifically, we setup the client and AP (with antennas) in a normal office
environment, and performed the association session for 100 times. A regular
wireless sniffer was used to collect traces during the experiment. We observed
that only 31 out of the 73 reachable states were reached during verifying these
sniffer traces. In particular, among the 40 states that require packet losses in
their jamming policy, only 6 states were reached in the 100 runs. The results
shows that the \analyzer{} and \jammer{} together can improve state coverage and
also provide validation reproducibility.

\subsubsection{Possible Violations Found}
\label{subsubsec:vio}

We now describe several implementation issues that cause the sniffer trace to be
rejected. We have reported our findings to the maintainers of corresponding
software module, and briefly summarize them here.

\noindent\textbf{Association Without Authentication.} We found if the first
\authresp{} packet was received by the client but the acknowledgment and all
following retransmissions of \authresp{} were jammed, the AP still accepted the
following \assocreq{} packet from the client. We originally thought the AP
should have rejected the association request since the \authresp{} packet
failed, thus authentication failed as well. After communicating with the
maintainer of \texttt{hostapd}, we realized this corner case was covered in the
latest 802.11 2016 standard, which states the AP considers the client as
\textit{authenticated} as soon as receiving the \authreq{} packet from the
client. We were previously referring to the 802.11 2012 standard, which did not
describe this case clearly.

\noindent\textbf{Double Association.} When the acknowledgment of the \assocreq{}
packet and all of its retransmissions were jammed, the client sent a new
\assocreq{} despite having received the \assocresp{} from the AP. This is
because from the client's point of view, the \assocreq{} packet failed, thus it
restarted the association process. However, the reception of the \assocresp{}
packet indicates that the AP actually received the \assocreq{} packet, thus
should have shortcut client's next association attempt.

\noindent\textbf{802.11X Deadlock.} When the acknowledgment of the first
\assocresp{} and all its retransmissions were jammed, the AP claimed the
association step failed and would not continue to the 802.11X authentication
stage.  However, the client actually received the \assocresp{} packet and was
waiting for the AP to initiated the 802.11 authentication step. Thus the client
and the AP entered a deadlock state.
  \section{Related Work}
\label{sec:related}

\noindent\textbf{Industry practice:}
Wireless protocols are typically tested both in a conducted setup and over-the-air.
Conducted tests provide a controlled environment. However, even for a conducted setup, we
are unaware of any commercially available attenuators that can selectively
trigger packet drops. This was corroborated in our conversations with \wifi{} chip
companies. The latency of existing variable attenuators is in the order of 10s of
milliseconds, while we need micro-second
latencies to trigger single packet drop, as achieved by {\jammer}. Over-the-air tests are
run to stress the entire RF subsystem including the antenna in a realistic
setup. However, these tests are left running for days to get the appropriate
test scenarios to be triggered, and the test time can be significantly
shortened by {\jammer}.

\noindent\textbf{Protocol verification:}
There has been work on model checking network protocol implementations~\cite{musuvathi2002cmc, godefroid1997model}.
The problem is different when trying to verify wireless systems, because (i) it
is non-trivial to trigger all possible states in wireless systems, and (ii) there
is uncertainty in the output captured by the wireless sniffer.  Wireless
sniffers have been widely used to analyze MAC behavior in enterprise wireless
networks~\cite{sheng:wicom2008,tan:tmc2014,yeo-wise04,yeo:witmemo2005,
Cheng:2006:JSP:1159913.1159920, Mahajan:2006:AMB:1159913.1159923}. However, this
body of research assumes correctness of the protocol implementation while
finding anomalies, while {\verifi} uses the sniffers to verify the
implementation.

\noindent\textbf{Model-based testing and validation:}
Model-based testing (MBT) is the approach of generating test cases by exploring
the model of a system under test~\cite{utting-mbt}.  Applications of MBT differ
in the modeling approach (UML, Statechart, first-order logic), the use of
test-case generation strategies (model checking, theorem proving) and the
application domains (automotive, distributed systems etc.).  One can view our
approach as the first instance of the approach where we use model
checker to explore the model to construct jamming policy that is used in
conjunction with the {\jammer} to drive the wireless system into interesting
states.  Our trace validation problem is also more complex due to the presence
of uncertainty in observations.  Similar problem happens when validating sampled
traces in the context of runtime
verification~\cite{bonakdarpour2011sampling,hauswirth2004low,arnold2008qvm,fei2006artemis}.
We incorporate the sniffer trace uncertainty problem~\cite{shi-rv16} but our new formulation allows leveraging any model checker instead of a customized checker (as in ~\cite{shi-rv16}).

 \section{Conclusions and Future Work}
\label{sec:conclusion}

Testing of wireless protocols is typically cumbersome and incomplete. These
systems are left running for several hours to ensure that the system does not
exhibit abnormal behavior. This takes a long time, and often bugs are not found
until after the product is released. In this paper we present {\verifi}, a new
automatic framework for validating wireless protocol implementation.  We have
described the design of \verifi{}, and provided implementations based on
\uppaal{} model checker and Software Defined Radio platforms (USRP and SORA). We
have evaluated our system via both micro benchmarks and an end-to-end cast study
on the \wifi{} link setup protocol, and showed that \verifi{} can improve
validation coverage and reproducibility.

We believe we have only scratched the surface in using
formal methods to validate wireless protocol implementations. Our techniques
could be further used for other applications as well, such as testing
the security aspects of wireless protocols. Moving forward, we are working in two
directions in the near term. First, we are extending the approach to detect
timing errors. Second, we are extending {\verifi} such that it can work for
other wireless protocols, such as a frequency-hopping Bluetooth, or a FDD
cellular system.

\clearpage
\newpage

{\footnotesize
    \bibliographystyle{abbrv}

}

\end{document}